\documentstyle[prl,aps,epsfig,multicol,ifthen,color]{revtex}

\input{epsf}
\begin{document}
\draft

\title{Laser cooling with electromagnetically induced
transparency: \\ Application to trapped samples of ions or neutral atoms}
\author{F.~Schmidt-Kaler, J.~Eschner,
G.~Morigi$^{(1)}$, C.~F.~Roos$^+$, D.~Leibfried$^*$, A.~Mundt, R.~Blatt}
\address{Institut f\"ur Experimentalphysik, University of Innsbruck,
A-6020 Innsbruck, Austria \\ $^{(1)}$Max-Planck-Institut f\"ur Quantenoptik,
D-85748 Garching, Germany} \maketitle

\begin{abstract}
A novel method of ground state laser cooling of trapped atoms utilizes the
absorption profile of a three (or multi-) level system which is tailored by a
quantum interference. With cooling rates comparable to conventional sideband
cooling, lower final temperatures may be achieved. The method was
experimentally implemented to cool a single Ca$^+$ ion to its vibrational
ground state. Since a broad band of vibrational frequencies can be cooled
simultaneously, the technique will be particularly useful for the cooling of
larger ion strings, thereby being of great practical importance for
initializing a quantum register based on trapped ions. We also discuss its
application to different level schemes and for ground state cooling of
neutral atoms trapped by a far detuned standing wave laser field.
\end{abstract}

\pacs{PACS: }

\begin{multicols}{2}

\section{Introduction}

The emergence of schemes that utilize trapped ions or atoms for
quantum information, and the interest in quantum statistics of
ultra cold atoms, have provided renewed interest for laser cooling
techniques. Starting with the case of a single atom, the present
goal is now to laser cool larger numbers of atoms and to prepare
them for applications e.g. as a quantum register. The cooling of a
large number of particles using lasers is a prerequisite for
coherent control of atomic systems.

Trapped ions in linear Paul traps are currently considered to be promising
candidates for a scalable implementation of quantum computation
\cite{ALMAGRO}. Long lived internal states serve to hold the quantum
information (qubits) and are manipulated coherently (single-bit gates) by
laser beams focused on the ions \cite{NAGERL99}. The excitation of the ion's
common vibrational motion (gate mode) provides the coupling between qubits
which is necessary for two-bit quantum gate between ions. The Cirac-Zoller
proposal for a two-bit gate \cite{CIRAC95} requires that initially the gate
mode is optically cooled to the ground state. Later M\"olmer and S\"orensen
proposed a two-bit gate which releases this condition and only requires
cooling into the Lamb-Dicke regime, typically with a mean vibrational quantum
number $\bar{m}\leq$1 \cite{MOELMER99,MOELMER00}.

In general, the motion of a string of $N$ cold trapped ions in a linear Paul
trap is described by $3N$ harmonic modes of vibration. Cooling {\em all
modes} close to the vibrational ground state is a prerequisite to realizing
quantum gates with high fidelity. In typical experimental situations this
requires cooling techniques which lead to lower mean vibrational state than
simple Doppler cooling. Sideband cooling has been shown to achieve
sub-Doppler temperatures \cite{Diedrich,MONROE95,KING98,ROHDE01} but proves
increasingly inefficient for cooling many modes. The recently proposed method
of EIT-cooling provides parallel multi-mode cooling while its experimental
effort is significantly reduced.

The first part of this paper (Sec.~\ref{EITprinc}) describes the principle of
the EIT-cooling method and illustrates its results with several realistic
examples, the cooling of a single Hg$^+$ ion held in a Paul trap, and of
neutral Rb atoms confined in a far-detuned optical dipole trap
(sect.~\ref{HgRb}). In section~\ref{expt} we report the first experimental
demonstration of the EIT-method, achieving ground state cooling of a single
trapped Ca$^+$ ion. The detailed description of all necessary experimental
ingredients and measurements is meant to provide a recipe for implementing
EIT-cooling in various experimental situations. We conclude by a proposal for
cooling a linear string of ions (Sec.~\ref{linear}), and discuss the
relevance for quantum information processing.

\section{The principle of EIT-cooling}

\subsection{Cooling basics}\label{Cbasics}

The theory of laser cooling of atoms in a harmonic trap has been extensively
discussed in the literature, see \cite{STENHOLM86} for a comprehensive
review. In many experimental situations a simple, intuitive rate equation
picture \cite{NEUHAUSER78} applies which can also be derived from the general
theory in the appropriate limits \cite{STENHOLM86}. In this section we recall
those intuitive considerations, a detailed theoretical derivation being
beyond the scope of this paper.

\begin{center}
\begin{figure}[tbp]
\epsfig{file=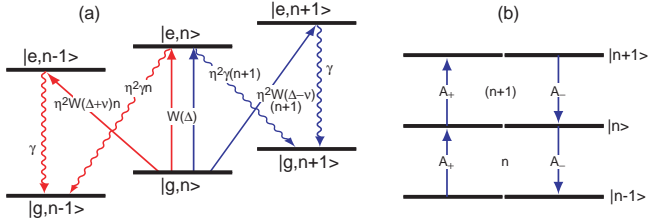,width=0.99\hsize} \vspace{\baselineskip}
\caption{(a) Cooling ($|n\rangle \to |n-1\rangle$) and heating
($|n\rangle \to |n+1\rangle$) transitions starting from state
$|g,n\rangle$, at lowest order in $\eta$. The relative
probabilites of the individual processes are indicated. (b)
Illustration of the rate coefficients for cooling and heating,
after \protect\cite{STENHOLM86}. \label{basics}}
\end{figure}
\end{center}

Consider a trapped 2-level atom with states $|g\rangle$ and $|e\rangle$,
excited below saturation by a laser detuned from resonance by $\Delta$. The
scattering rate, i.e. steady-state excited state population times decay rate,
is denoted by $W(\Delta)$. The atom is coupled to one vibrational degree of
freedom, a harmonic oscillator with frequency $\nu$ and quantum states
$|n\rangle$, such that the internal transition $|g\rangle \leftrightarrow
|e\rangle$ splits up into transitions $|g,n\rangle \leftrightarrow
|e,n'\rangle$. The coupling between different motional states is mediated by
the recoil of a laser or spontaneous photon. We restrict ourselves to the
so-called Lamb-Dicke regime $\eta \equiv k\sqrt{\langle x^2 \rangle} \ll 1$
where the spatial extension of the motional wave packet $\sqrt{\langle x^2
\rangle}$ is smaller than the laser wavelength $\lambda=2\pi/k$. In this
limit the relevant transitions are $|n\rangle \to |n\rangle$ (carrier) and
$|n\rangle \to |n\pm1\rangle$ (sidebands). See Fig.~\ref{basics}a for an
illustration.

The probability that scattering of a photon takes the atom from $|g,n\rangle$
to $|g,n\pm1\rangle$ has two contributions, corresponding to the two possible
intermediate states $|e,n\rangle$ or $|e,n\pm1\rangle$. The rate coefficients
$A_{\pm}$ for cooling and heating transitions (see Fig.~\ref{basics}b) are
therefore given by
\begin{equation}
\label{Aplusminus} A_{\pm} = \eta^2 \left( W(\Delta) + W(\Delta\mp\nu)
\right)~.
\end{equation}
Summing up all processes in lowest order $\eta$, a rate equation for the
populations $p_n(t)$ of the states $|n\rangle$ is obtained which leads to a
simple dynamical equation for the mean vibrational excitation $\langle n(t)
\rangle = \sum n p_n(t)$,
\begin{equation}
\label{Evolution} \frac{d}{dt}\langle n \rangle = - (A_- - A_+) \langle n
\rangle + A_+~.
\end{equation}

In the more general case that the laser $k$ vektor is at an angle $\theta$ to
the direction of vibration, and that the spontaneous recoils are spatially
distributed with an average projection $\alpha$ on the vibrational axes
\cite{noteonalpha}, Eq.~(\ref{Aplusminus}) changes into
\begin{equation}
\label{Aplusminus1} A_{\pm} = \eta^2 \left( \alpha W(\Delta) + \cos^2(\theta)
W(\Delta\mp\nu) \right)~.
\end{equation}
Using Eqs.~(\ref{Evolution},\ref{Aplusminus1}), the cooling limit $\bar{m}
\equiv \langle n(\infty) \rangle = A_+/(A_--A_+)$ and the cooling time
constant, $\tau_{cool}=(A_--A_+)^{-1}$ can be easily evaluated. One finds the
well-known cooling limits $\bar{m}\nu \simeq \gamma/2$ for Doppler cooling
where $\gamma>\nu$ and $\Delta=-\gamma/2$, and $\bar{m} \ll 1$ for sideband
cooling where $\gamma<\nu$ and $\Delta=-\nu$ \cite{STENHOLM86}.

It is useful to look at the transitions displayed in Fig.~\ref{basics}a from
a different viewpoint: Absorption on one of the sidebands causes a drift of
the motional energy while carrier absorption, being followed by either
cooling or heating emission, causes a diffusion of the energy which in
average contributes to heating. Therefore for optimum cooling, absorption on
the cooling sideband should be maximized while carrier and heating-sideband
absorption should be minimized. This leads to the principle of EIT-cooling.

\subsection{EIT-cooling principle} \label{EITprinc}

Electromagnetically induced transparency (EIT), like coherent population
trapping or "dark resonance", is a manifestation of quantum interference
between atomic transition amplitudes. A review can be found in
\cite{HARRIS97}. EIT arises in three- (or multi-) level systems and consists
in the cancellation of the absorption on one transition induced by
simultaneous coherent driving of another transition. Consider a 3-level atom
as shown in Fig.~1, with ground state $|g\rangle$, stable or metastable state
$|r\rangle$ and excited state $|e\rangle$. State $|e\rangle$ has linewidth
$\gamma$ and is coupled to $|g\rangle$ and $|r\rangle$ by dipole transitions
which are laser-excited. EIT arises when the detunings of the two lasers from
state $|e\rangle$ are equal: The system evolves into a coherent superposition
of $|g\rangle$ and $|r\rangle$, and light scattering seizes.

In order to use this situation for cooling, the transition $|r\rangle \to
|e\rangle$ is excited by an intense ("coupling") laser ($\Omega_r \sim
\gamma$) at detuning $\Delta_r$ above resonance. Then the absorption spectrum
on the transition $|g\rangle \to |e\rangle$ seen by the other ("cooling")
laser is described by a Fano-like profile \cite{COHEN92} which has three
characteristic features, see Fig.~1a,b: the broad resonance at $\Delta_g
\simeq 0$, the dark resonance (EIT) at $\Delta_g = \Delta_r$, and the narrow
resonance at $\Delta_g = \Delta_r+\delta$, where
\begin{equation}
\label{ACStark} \delta=(\sqrt{\Delta_r^2+\Omega_r^2}-|\Delta_r|)/2
\end{equation}
is the ac-Stark shift created by the coupling laser.
\begin{center}
\begin{figure}[tbp]
\epsfig{file=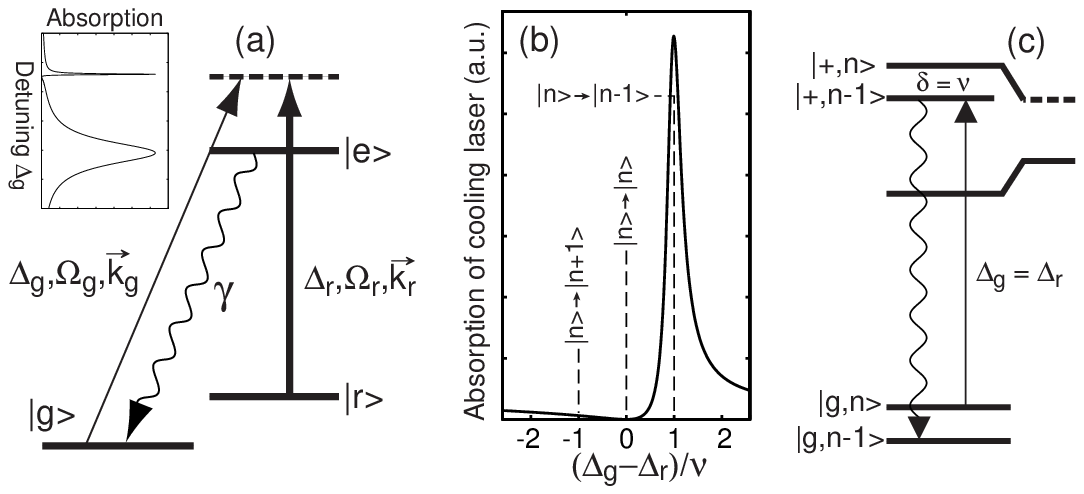,width=0.99\hsize} \vspace{\baselineskip}
\caption{(a) Levels and transitions of the EIT-cooling scheme. The
inset shows schematically the absorption rate on $|g\rangle\to
|e\rangle$ when the atom is strongly excited above resonance on
$|r\rangle \to |e\rangle$. (b) Absorption of cooling laser around
$\Delta_g = \Delta_r$ (solid line); probabilities of carrier
($|n\rangle \to |n\rangle$) and sideband ($|n\rangle \to |n \pm
1\rangle$) transitions when $\Delta_g = \Delta_r$ are marked by
dashed lines. (c) Dressed state picture: the cooling laser excites
resonantly transitions from $|g,n\rangle$ to the narrow dressed
state denoted by $|+,n-1\rangle$ which preferentially decays into
$|g,n-1\rangle$. \label{EITprinciple}}
\end{figure}
\end{center}

For EIT-cooling, the laser frequencies are set to the dark resonance
condition, $\Delta_g = \Delta_r$. Then, taking into account the harmonic
motion, all $|g,n\rangle \to |e,n\rangle$ transitions are cancelled.
Furthermore, by choosing a suitable Rabi frequency $\Omega_r$, the spectrum
is designed such that the $|g,n\rangle \to |e,n-1\rangle$ (red) sideband
corresponds to the maximum of the narrow resonance, whereas the blue sideband
falls into the region of the spectrum of small excitation probability, as
shown in Fig.~1b. The conditions for enhancing the red-sideband absorption
while eliminating the carrier is therefore:
\begin{equation}
\label{settings} \Delta_g = \Delta_r \hspace{5mm} ; \hspace{5mm} \delta
\simeq \nu
\end{equation}

In summary, due to the EIT condition no absorption happens unless the ion is
moving, and by adjusting the coupling Rabi frequency such that the bright
resonance matches the red sideband, absorption accompanied by the loss of one
phonon is made much more probable than absorption accompanied by gaining one
phonon. The fact that the diffusion normally caused by $|n\rangle \rightarrow
|n\rangle$ absorption is eliminated is most important for three-dimensional
cooling, since the cooling limit $\langle n \rangle$ becomes independent of
the projection of the vibrational mode on the laser direction
\cite{MORIGI00}.

\subsection{Cooling dynamics}

\begin{center}
\begin{figure}[tbp]
\epsfig{file=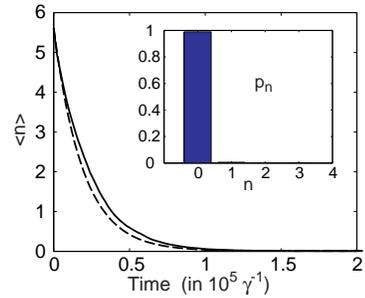,width=0.55\hsize}
\vspace{\baselineskip} \caption{Onset: $\langle n\rangle$ as a
function of time calculated with full Monte Carlo simulation
(solid line) and rate equation (dashed line). Parameters are
$\Omega_r=\gamma$, $\Omega_g=\gamma/20$, $\nu=\gamma/10$,
$\eta=0.145$, $\Delta_g=\Delta_r=2.5\gamma$. Inset: Steady state
distribution $p_n$. \label{3levelsim}}
\end{figure}
\end{center}

A calculation starting from the full quantum mechanical master equation
\cite{Giovanna} shows that in the Lamb-Dicke regime and below saturation of
the cooling transition, the EIT-cooling process can be approximated by the
same rate equation discussed above, only that now the appropriate
coefficients $A_{\pm}$ contain the quantum interference around $\Delta_{\pi}
= \Delta_{\sigma}$. The explicit form of the coefficients is given in
\cite{MORIGI00}. Fig.~\ref{3levelsim} shows the cooling dynamics calculated
from a Monte-Carlo simulation of the master equation and from the rate
equation approximation.

\subsection{Example calculations} \label{HgRb}

EIT-cooling shows to be a robust technique even if the system involved
deviates from the ideal 3-level scheme. For example, in the experiments
described below, due to non-ideal polarizations of the laser beams, photons
were scattered from a fourth level which introduced extra heating. The
theoretical cooling limit for this particular case increases by about a
factor of 3, and the experiment still yielded cooling close the ground state
\cite{ROOS00}, see Sec.~\ref{expt}.

\begin{center}
\begin{figure}[tbp]
\epsfig{file=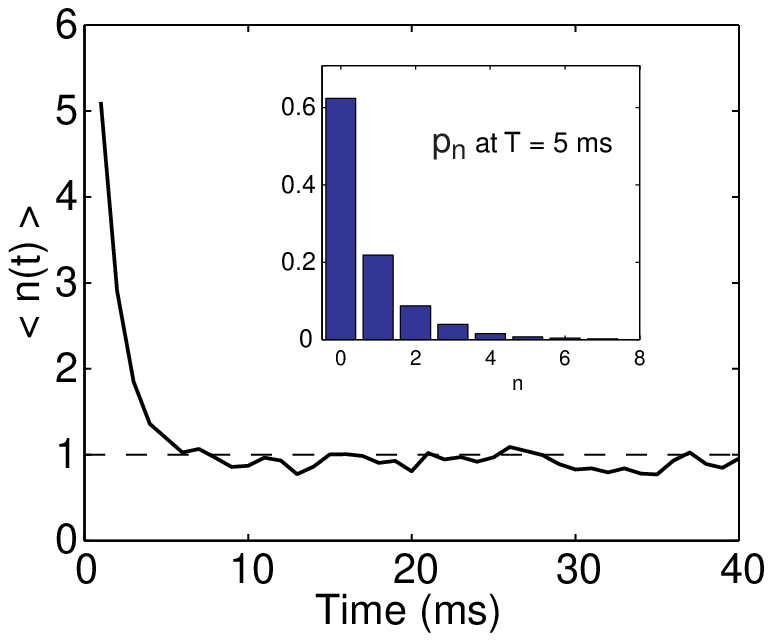,width=0.55\hsize} \vspace{\baselineskip}
\caption{Onset: $\langle n\rangle$ as a function of time
calculated with full Monte Carlo simulation for a single trapped
Hg$^+$ ion. Parameters are $\gamma=64$~MHz, $\nu=1.5$~MHz,
$\Omega_r=21$~MHz, $\Omega_g=4$~MHz, $\Delta_r=\Delta_g=80$~MHz,
$\Omega_{repump}=2$~MHz, $\Delta_{repump}=0$. Inset: Steady state
distribution $p_n$. \label{Hgsim}}
\end{figure}
\end{center}

Another interesting situation is when the excited state decays through an
additional channel into a fourth level from which it has to be repumped with
an additional laser. We studied the example of a single trapped Hg$^+$ ion
where the states $|g\rangle,|r\rangle,|e\rangle$ for EIT-cooling are taken to
be the ${\rm S}_{1/2}(F=1,m_F=1,0)$ and ${\rm P}_{1/2}(F'=1,m_F=1)$ levels,
respectively. The ${\rm S}_{1/2}(F=0,m_F=0)$ level into which $|e\rangle$
decays with 1/3 probability is pumped out resonantly back to $|e\rangle$. Our
model calculation is shown in Fig.~\ref{Hgsim}. A cooling limit of $\langle n
\rangle \simeq 1$ is achieved which is significantly below the Doppler
cooling limit of about 20 and which could help to improve the Hg$^+$ single
ion frequency standard \cite{RAFAC00}.

\begin{center}
\begin{figure}[tbp]
\epsfig{file=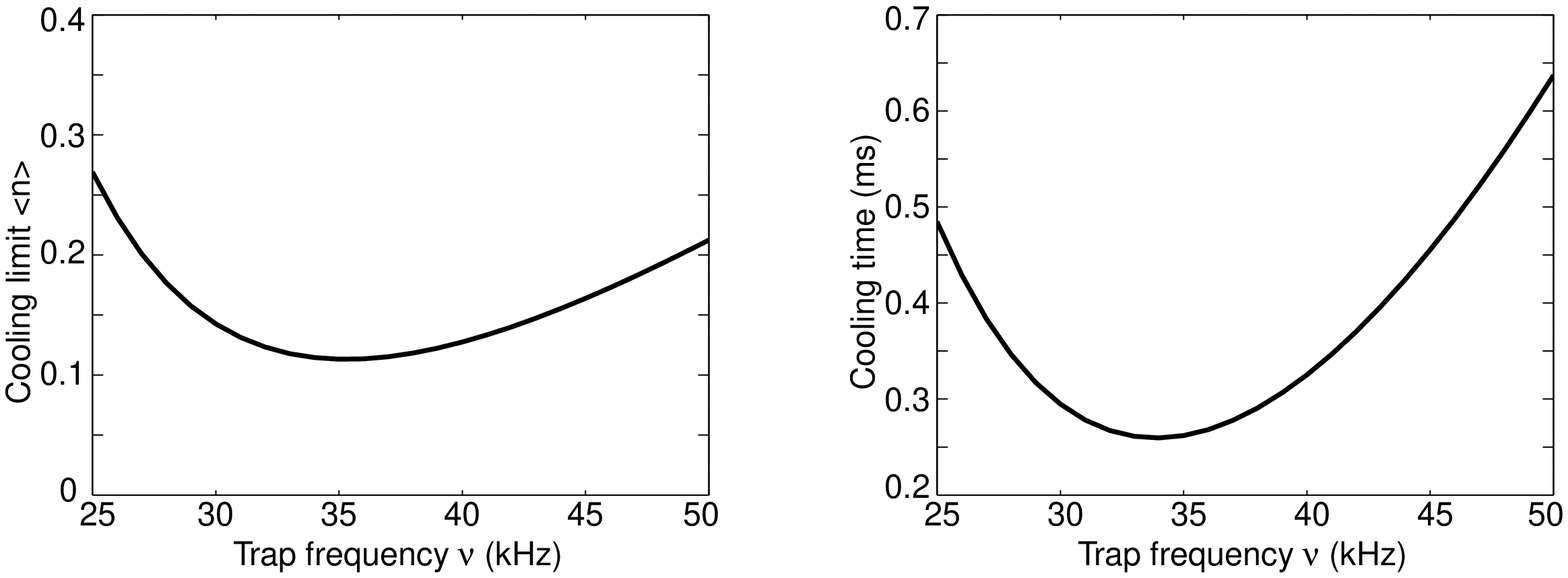,width=0.99\hsize} \vspace{\baselineskip}
\caption{Estimate for EIT-cooling of $^{87}$Rb atoms in a
CO$_2$-laser standing wave trap: cooling limit (left) and cooling
time (right) for trap frequencies of 25~..~50~kHz. Parameters are
$\gamma=6$~MHz, $\Omega_r=1.2$~MHz, $\Omega_g=0.1$~MHz,
$\Delta_r=\Delta_g=10$~MHz, $\Omega_{repump}=1$~MHz,
$\Delta_{repump}=0$. \label{dipole}}
\end{figure}
\end{center}

Finally, it would certainly be also interesting to apply EIT-cooling in
cold-atom experiments, e.g. to Rb or Cs. In fact, the outer $m$-states of any
$J\to J$ transition offer a suitable level systems for its implementation.
Specifically, we give an estimate for the application of EIT-cooling to
$^{87}$Rb atoms in a CO$_2$-laser standing wave dipole trap \cite{WEITZ}. One
main difference to an ion trap are that the trap frequency is generally lower
and varies along the laser beam waist. We assumed a distribution of trap
frequencies between 25 and 50~kHz, corresponding to the intensity variation
over one Rayleigh length. We applied the rate equation model to the
$(F=1,m_F=0,1) - (F'=1,m_F=1)$ 3-level system and included resonant repumping
from $F=2$. The detailed parameters and the calculated results are displayed
in Fig.~\ref{dipole}. According to that estimate it should be possible to
cool to $\langle n \rangle<1$, much below the Doppler cooling limit, on a
time scale of ms.

It should be remarked that in this case the motional state before cooling may
be outside the Lamb-Dicke regime, such that initially the considerations of
Sec.~\ref{Cbasics} do not apply. Instead, higher order sidebands, i.e.
transitions with $|n-n'|>1$ have to be taken into account. In the example of
Fig.~\ref{dipole} one finds that the absorption probability for cooling
transitions is larger than that for heating transitions for many higher-order
sidebands (over several 100~kHz), therefore starting outside the Lamb-Dicke
regime does not seem to restrict the application of EIT-cooling.

\section{EIT-cooling of a single Ca$^+$ ion to the ground state} \label{expt}

\subsection{Levels and transitions in Ca$^+$}

\begin{center}
\begin{figure}[t]
\epsfig{file=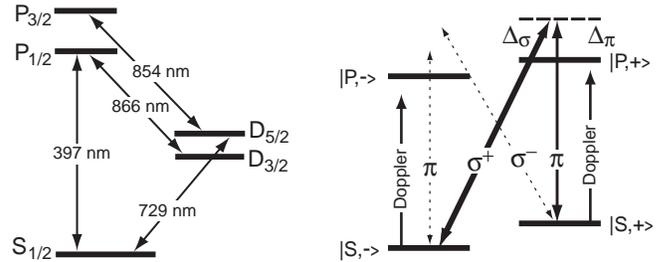,width=0.99\hsize}\vspace{\baselineskip}
\caption{\label{levels}Levels and transitions in $^{40}$Ca$^+$
used in the experiment (left side). The S$_{1/2}$ to P$_{1/2}$
transition is used for Doppler cooling and for EIT-cooling, and
the scattered photons are observed to detect the ion's quantum
state. The narrow S$_{1/2}$ to D$_{5/2}$ transition serves to
investigate the vibrational state (see section \ref{t}). Shown on
the right are the Zeeman sublevels of the S$_{1/2}$ and P$_{1/2}$
states (denoted by $|{\rm S},\pm\rangle$ and $|{\rm
P},\pm\rangle$) and the laser frequencies relevant for cooling.
The solid lines labeled by $\pi$ and $\sigma^+$are used for
EIT-cooling, with $\Delta_{\pi} = \Delta_{\sigma}$. Dashed lines
indicate further transitions which involve the $|{\rm P},-\rangle$
level. }
\end{figure}
\end{center}

The theoretical background of the method described above has to be
modified only slightly to apply it to our experiment with a single
trapped Ca$^+$ ion. We implemented the EIT-scheme on the ${\rm
S}_{1/2}\to {\rm P}_{1/2}$ transition, whose Zeeman sublevels
$m=\pm1/2$ constitute a four-level system. We denote these levels
by $|{\rm S},\pm\rangle$ and $|{\rm P},\pm\rangle$, see
Fig.~\ref{levels}. Three of the levels, $|{\rm S},\pm\rangle$ and
$|{\rm P},+\rangle$, together with the $\sigma^+$- and
$\pi$-polarized laser beams, form an effective three-level system
of the kind considered above \cite{MORIGI00}. The modifications
due to the fourth level $|{\rm P},-\rangle$ are discussed in
Sec.~\ref{procedure}.

\subsection{Experimental setup}

\begin{center}
\begin{figure}[t]
\epsfig{file=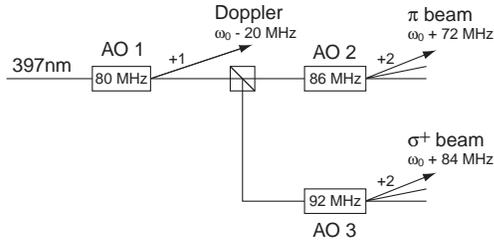,width=0.75\hsize} \vspace{\baselineskip}
\caption{\label{setup}Optical setup for the light beams at 397~nm.
Inset: Scheme of the frequencies, as used for the Doppler- and for
the EIT-cooling. Both EIT-beams are blue detuned. The Doppler
cooling beam is red from Ca$^+$ ion resonances $|-,S\rangle
\rightarrow |-,P\rangle$ and $|+,S\rangle \rightarrow
|+,P\rangle$.}
\end{figure}
\end{center}

The Ca$^+$ system seems ideal for the demonstration of the
EIT-cooling method since in previous experiments we have found
very small heating rates \cite{ROOS99} and therefore the cooling
results can be measured with high precision. We will briefly
introduce the experimental setup, for details we refer to
\cite{ROOS99,ExpIonsIbk}. First we describe all light sources
which are necessary for the experiment:

For the S$_{1/2} \to $P$_{1/2}$ dipole transition at 397~nm, light from a
Ti:Sapphire laser near 793~nm is frequency doubled in a LBO crystal inside a
build-up cavity. Up to 5~mW of light near 397~nm are sent to the trap setup
through a single mode UV fiber. As displayed in Fig.~\ref{setup}, we use the
beam deflected into $+1^{st}$ order from an acousto-optical modulator (AO~1)
driven at 80~MHz for Doppler cooling. Typically, this radiation is
red-detuned from the resonance by 20~MHz. The two beams for EIT-cooling,
labeled $\sigma^+$ and $\pi$, are derived from the UV fiber output as the
$+2^{nd}$ deflection orders of two further acousto-opical modulators AO~2 and
AO~3 driven at 86~MHz and 92~MHz, respectively. We chose a detuning of
$\Delta_{\sigma} = \Delta_{\pi} \simeq $ 75~MHz (natural linewidth of the
transition is 20~MHz). The difference in the AO~2 and 3 drive frequencies
compensates for the 12~MHz Zeeman splitting of the $|{\rm S},\pm\rangle$
states in the quantization B-field of 4.4~Gauss. The three output beams are
focused onto the single ion in the Paul trap from different directions
corresponding to their respective polarisations and the orientation of the
B-field. For a reproducible and fine tuning of the light intensity (typically
few 10~$\mu$W in a waist of $\simeq$ 60 $\mu$m) the rf-drive power of all
AO's is computer-controlled. The rf-power, and therefore the light beams, can
also be switched on and off within a few $\mu$s.

Light from a second Ti:Sapphire laser near 729~nm is used for excitation on
the quadrupole transition from the S$_{1/2}$ ground state to the metastable
D$_{5/2}$ state (spontaneous lifetime 1~s). The laser frequency is stabilized
to a bandwidth of $\delta \nu \leq$ 100~Hz using a stable high-finesse
optical reference cavity and a Pound-Drever-Hall servo circuit. The laser
frequency can be tuned by an acousto-optical modulator, its intensity is
controlled through the rf-drive power. With this laser we can selectively
excite vibrational sidebands of the S$_{1/2} - $D$_{5/2}$ transition at a few
MHz detuning from the carrier. The relative excitation probability on the
first red and the first blue sideband reveals information about the ion's
vibrational quantum state as described in sect.~\ref{t}.

The light of grating stabilized laser diodes near 866~nm and 854~nm serves to
pump the ion out of both metastable D levels. While the light of the laser at
866~nm is turned on continuously troughout the experiment, the laser at
854~nm is also switched on and off by a computer-controlled acousto-optical
modulator.

Our ion trap is a 3D-quadrupole Paul trap: A ring with inner
diameter of 1.4~mm is formed from a Mb-wire of 0.2~mm diameter,
and two endcaps are made out of sharpened tips of the same
material at 1.2~mm distance \cite{ROOS99}. The ring electrode is
driven by $\simeq$ 650~V radio frequency at 20.9~MHz, while the
tips are held near ground potential. For the experiments described
below, a single $^{40}$Ca$^+$ ion is generated by electron
bombardment of a weak thermal Calcium beam. Its oscillation
frequencies in the trap potential ($\nu_x$, $\nu_y$, $\nu_z$) are
(1.69, 1.62, 3.32)~MHz. Additional compensation electrodes,
together with a differential voltage applied to the tips, can be
used to compensate for stray static electric fields and to shift
the ion into the center of the rf-potential.

\subsection{Measuring the vibrational quantum state} \label{t}

\begin{center}
\begin{figure}[t]
\epsfig{file=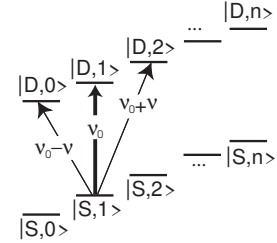,width=0.4\hsize}\vspace{\baselineskip}
\caption{Principle of temperature measurement. Transitions from
$|{\rm S},n\rangle$ to $|{\rm D},n\pm1\rangle$ are selectively
laser-excited and the relative excitation probability is measured.
$\nu_0$ is the bare atomic transition frequency. \label{dressed}}
\end{figure}
\end{center}

The average motional quantum number $\bar{m}$ can be measured with high
precision using laser excitation of vibrational sidebands, which are
spectrally well resolved since the linewidth of the transition, here S$_{1/2}
- $D$_{5/2}$, is well below the vibrational frequencies of the trapped ion.
The situation is schematically depicted in Fig.~\ref{dressed}, where the
level scheme obtained for a two-level atom in a harmonic trapping potential
is shown together with the carrier ($|n\rangle - |n\rangle$) and the first
sideband ($|n\rangle - |n\pm 1\rangle$) transitions.

The excitation probabilities $P_{red}$ and $P_{blue}$ on the $|S,n\rangle$ to
$|D,n-1\rangle$ transition (red sideband) and the $|S,n\rangle$ to
$|D,n+1\rangle$ transition (blue sideband) are proportional to $n$ and $n+1$,
respectively. $P_{red}$ drops to zero for $|n=0\rangle$, see
Fig.~\ref{dressed}. For a thermal phonon distribution with mean phonon number
$\bar{m}$, the probability for finding $n$ phonons is $p_n(\bar{m}) =
\bar{m}^n/ (\bar{m}+1)^n$. Thus $\bar{m}$ is directly related to the measured
ratio $P_{red} / P_{blue} = \bar{m} / (\bar{m}+1)$. It can be shown that this
result holds for incoherent {\em and} coherent excitation if the phonon
distribution $p_n(\bar{m})$ is thermal. Our cooling results will be specified
either as $\bar{m}$ or by the zero-phonon occupation $p_0$ for the three trap
vibrational modes.

One way to determine the cooling result is to first excite a vibrational
sideband at 729~nm and then probe the S$_{1/2} - $P$_{1/2}$ transition at
397~nm (shelving method) \cite{Diedrich}. If sideband excitation to the
D$_{5/2}$ state was successful, no fluorescence is emitted on the S$_{1/2} -
$D$_{5/2}$ transition, whereas in the other case we observe many scattered
photons. This procedure is repeated 100 times on both the red and the blue
sideband, and the difference in their excitation probabilities yields
$\bar{m}$ and $p_0$.

The second method we used is driving coherent Rabi oscillations on the blue
sideband \cite{ROOS99,Meekhof}. We record the excitation probability
$P_{blue}(t)$ as a function of 729~nm excitation time $t$. Theoretically,
$P_{blue}(t) = \sum p_n(\bar{m}) \sin (\eta \Omega_{Rabi} \sqrt{n+1}~~t)$,
where $\Omega_{Rabi}$ denotes the S$_{1/2} - $D$_{5/2}$ Rabi frequency and
$\eta$ the Lamb-Dicke factor \cite{BLOCKLEY92}. The analysis of a time
evolution $P_{blue}(t)$ reveals the mean phonon number $\bar{m}$ or can even
be used to measure all coefficients of the phonon distribution
$p_n(\bar{m})$.

\subsection{Setting the power level for the EIT-beams} \label{ac}

Applying the EIT-cooling principle to our system, we find that the intensity
of the EIT-$\sigma^+$ beam should be such that the Stark shift equals the
frequency of the vibrational mode to be cooled. If the waist sizes of the
beams are known, their intensities can be calculated. There are also a few
possible methods to experimentally determine the ac-Stark shift or the
corresponding Rabi frequencies:
\\
a) A rough estimate may be gained from the fluorescence rate as a function of
light power. Finding first the approximate saturation power level, one can
then deduce the necessary laser power for both EIT-beams.
\\
b) In the $\{\rm{S}_{1/2},\rm{P}_{1/2},\rm{D}_{3/2}\}$ three-level
system excited by the 866~nm and 397~nm lasers, one can measure
the fluorescence rate if one of the laser frequencies is tuned
over the resonance and the other is kept fix. Optical Bloch
equations are then used to fit the excitation spectrum and
determine the relevant Rabi frequencies \cite{DarkRes}.
\\
c) The time constant of optical pumping between the $|{\rm S}, \pm\rangle$
states may be used. To probe their population, one of the two states must be
selectively excited to the D$_{5/2}$ level.
\\
d) One can measure the broadening and ac-Stark shift of the narrow S$_{1/2}$
to D$_{5/2}$ transition when the ion is simultaneously illuminated with a
pulse of 729~nm and one of the EIT-beams.

We used the last method to determine the ac-Stark shift $\delta$ as a
function of power in the EIT-beams. From this, we could extrapolate to the
power level where $\delta = \nu_{trap}$. In more detail, the ion was excited
on the ${\rm S}_{1/2} (m=1/2) \rightarrow {\rm D}_{5/2} (m=5/2)$ transition
with a 3~$\mu$s pulse at 729~nm and {\em simultaneously} with a
EIT-$\sigma^+$ pulse. Although the method suffered from asymetric excitation
line shapes and from rapid optical pumping out of the S$_{1/2}, m=1/2$ level,
it lead to a good estimate for the right power level. For a final
optimization of the EIT-$\sigma^+$ beam power we used the cooling results for
a specific vibrational mode \cite{ROOS00}.

\subsection{EIT-cooling experimental procedure} \label{procedure}

We investigated EIT cooling of the $y$ and the $z$ oscillation at 1.62 and
3.32~MHz, respectively. The experiment proceeded in three steps: Doppler
precooling, EIT-cooling and finally determination of the vibrational quantum
state, see Fig.~\ref{pulse}:

\begin{center}
\begin{figure}[t]
\epsfig{file=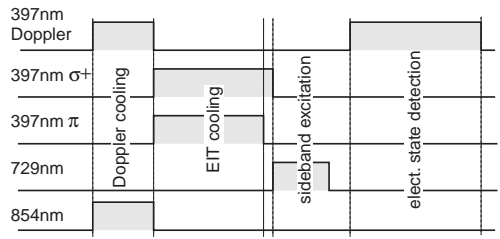,width=0.75\hsize}\vspace{\baselineskip}
\caption{\label{pulse} Experimental pulse sequence. A 1.5~ms
period of Doppler cooling is followed by a period of EIT-cooling
(varied between 0 and 7.9~ms). The $\sigma^+$-beam is left on
50~$\mu$s longer than the $\pi$-beam to optically pump the ion
into $|{\rm S}, +\rangle$. The analysis of the vibrational state
uses a pulse (of length $t$) at 729~nm for sideband excitation,
followed by a Doppler pulse at 397~nm to probe the shelving. The
laser at 866~nm continuously empties the D$_{3/2}$ level. The
whole sequence of 20~ms duration is repeated 100 times for each
sideband and the average excitation rate is calculated. To avoid
50~Hz magnetic field fluctuations the sequence is triggered by the
electric power line.}
\end{figure}
\end{center}

(i) We Doppler precooled the ion on the S$_{1/2}$ to P$_{1/2}$ transition at
397~nm (natural linewidth $\Gamma \simeq$~20~MHz). A detuning of
approximately -20~MHz with respect to the ${\rm S}_{1/2} - {\rm P}_{1/2}$
transition line was chosen. To avoid optical pumping into the D$_{3/2}$
states, we used the repumping beam near 866~nm \cite{ExpIonsIbk}. Doppler
cooling was applied for 1.5~ms. The theoretical Doppler cooling limit on this
transition of 0.5~mK corresponds to mean vibrational quantum numbers of
$\bar{m}^{z} \approx$ 3 and $\bar{m}^{x} \approx \bar{m}^{y} \approx$ 6. The
cooling limits reached in our experiment were higher, due to the fact that
the simple assumption of a two level system in the determination of the
Doppler limit does not hold in our case and because the 12~MHz Zeeman
splitting of the S$_{1/2}$ state does not permit optimum detuning for all
transitions between the substates. We experimentally determined the mean
excitation numbers after Doppler cooling to be $\bar{m}^z=6.5(1.0)$ and
$\bar{m}^y=16(2)$.

(ii) After Doppler cooling, we applied the two EIT-beams, 397nm-$\sigma^+$
and 397nm-$\pi$ for up to 7.9~ms. A 1.8~ms pulse duration was found to be
sufficient, longer cooling time did not lower the mean phonon number (see
Sec.~\ref{dyn} where we will discuss the cooling time). The power levels were
set as discussed above in Sec.~\ref{ac}. For our setup with laser beam waists
of $\simeq$~50~$\mu$m, a laser power of $\leq$~50~$\mu$W was used for the
$\sigma^+$-beam.
\\
The $k$-vectors of the two EIT-cooling beams enclosed an angle of $125^\circ$
and illuminated the ion in such a way that their difference $\Delta \vec{k}$
had a component along all trap axes ($(\phi_x,\phi_y,\phi_z) =
(66^\circ,71^\circ,31^\circ)$, where $\phi_i$ denotes the angle between
$\Delta \vec{k}$ and the respective trap axis). As a result, {\em all}
vibrational directions were cooled. Since the vacuum recipient did not allow
for two orthogonal beams, we could not realize the ideal constellation of
Fig.~\ref{levels}: While the $\sigma^+$-beam was parallel to the quantization
axis as desired, the $\pi$-beam was not orthogonal to it and therefore had a
$\sigma^-$-component. The extra excitation due to this non-ideal
constellation is indicated in Fig.~\ref{levels} by a dashed line. It results
in a slightly higher final mean phonon number. A detailed discussion is found
in Ref.~\cite{ROOS00}.

(iii) We analyzed the vibrational state after EIT-cooling by
spectroscopy on the ${\rm S}_{1/2} \to {\rm D}_{5/2}$ quadrupole
transition at 729~nm, as described in Sec.~\ref{t}. Finally a
diode laser at 854~nm served to repump the ion from the D$_{5/2}$
to the S$_{1/2}$ level.

\section{EIT-cooling results}
\subsection{Cooling results for a single mode of vibration}

\begin{center}
\begin{figure}[t]
\epsfig{file=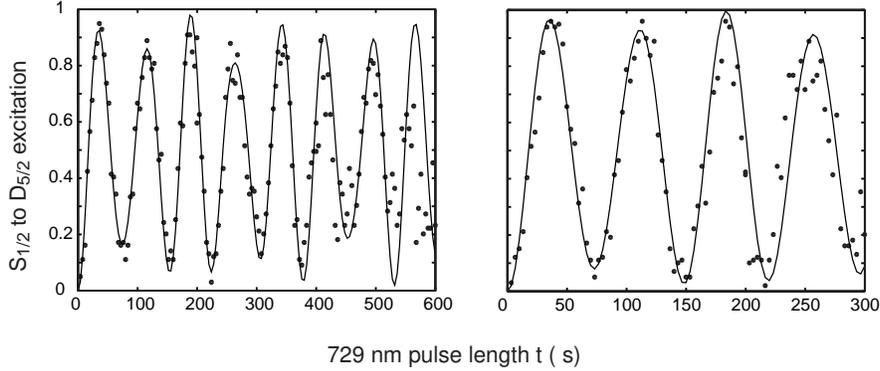,width=0.65\hsize}\vspace{\baselineskip}
\caption{\label{singlemode} Rabi oscillations excited on the blue
sideband of the radial y-mode (left) and the axial z-mode (right)
of a single ion. From the theoretical curves for $P_{blue}(\tau)$,
mean phonon numbers of $\bar{m}^y=0.18$ and $\bar{m}^z=0.1$ are
determined (see Sec.~\ref{t}).}
\end{figure}
\end{center}

With the ac-Stark shift $\delta$ set to the frequency of the
radial y-mode, we monitored the vibrational state after
EIT-cooling by exciting the blue sideband of the $|{\rm
S},+\rangle \to$ D$_{5/2}(m=+5/2)$ transition with a 729~nm pulse
and then measuring the $|{\rm S},+\rangle$ level occupation as a
function of the pulse length $t$ \cite{ROOS99,ROOS00}. The
observed Rabi-oscillations were subsequently fitted to determine
the mean vibrational occupation number $\bar{m}^y$ \cite{Meekhof},
see Fig.~\ref{singlemode}. The lowest mean vibrational number
$\bar{m}^y=0.18$ observed corresponds to a 84$\%$ ground state
probability. We repeated this experiment on the $z$-mode at
3.3~MHz after having increased the intensity of the
$\sigma^+$-beam to adjust $\delta$. For this mode, a minimum mean
vibrational number of $\bar{m}^z=0.1$ was obtained, corresponding
to a 90$\%$ ground state probability.
\end{multicols}
\begin{multicols}{2}

We found the cooling results largely independent of the intensity of the
$\pi$-beam as long as it is much smaller than the $\sigma^+$ intensity. In
our experiment the intensity ratio was $I_{\sigma}/I_\pi \simeq 100$ and we
varied the intensity of the $\pi$-beam by a factor of 4, with no observable
effect on the final $\bar{m}$.

\subsection{Cooling dynamics} \label{dyn}

\begin{center}
\begin{figure}[t]
\epsfig{file=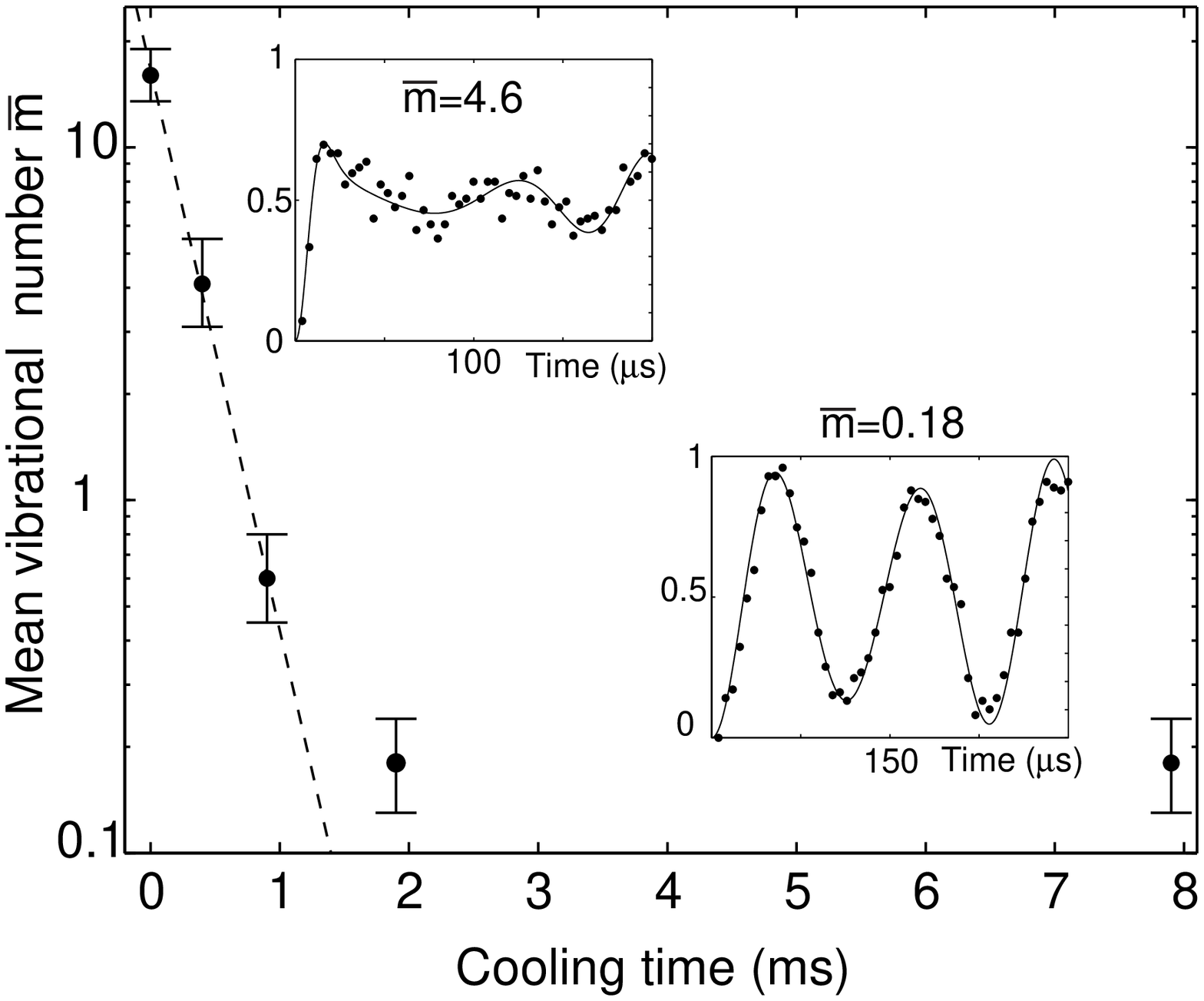,width=0.85\hsize}\vspace{\baselineskip}
\caption{\label{coolingdyn} Mean vibrational quantum number
$\bar{m}^y$ versus EIT cooling time. The insets show Rabi
oscillations excited on the upper motional sideband of the $|{\rm
S},+\rangle \to$ D$_{5/2}(m=+5/2)$ transition, after 0.4~ms (left)
and 7.9~ms (right) of EIT cooling. A thermal distribution is
fitted to the data to determine $\bar{m}^y$ in both cases.}
\end{figure}
\end{center}

By determining the dependence of the mean vibrational quantum number on the
EIT cooling time, we measured the cooling time constant for the y-mode to be
250~$\mu$s, as shown in Fig.~\ref{coolingdyn}.

\subsection{Cooling of two modes}

\begin{center}
\begin{figure}[t]
\epsfig{file=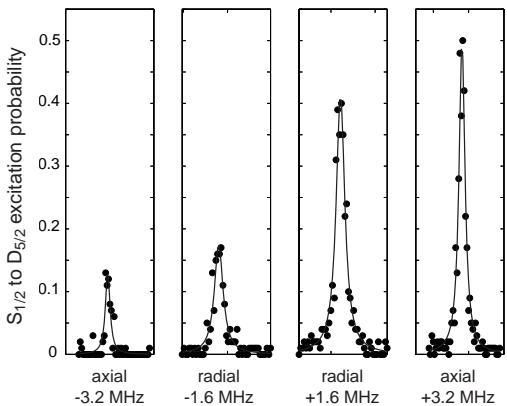,width=0.80\hsize}\vspace{\baselineskip}
\caption{EIT-cooling of two modes at 1.6~MHz and 3.2~MHz
simultaneously. From the sideband excitation rate after cooling we
deduce a ground state occupation number of 73$\%$ for the axial
mode (3.2~MHz) and 58$\%$ for the radial mode (1.6~MHz).
\label{twomodes}}
\end{figure}
\end{center}

To show that the EIT method is suitable to simultaneously cool several
vibrational modes with significantly different frequencies of oscillation, we
chose the axial $z$-mode at 3.3~MHz, and the radial $y$-mode at 1.62~MHz. The
intensity of the $\sigma^+$-beam was set such that the ac-Stark shift was
roughly halfway between the two mode frequencies. Again we applied the EIT
cooling beams for 7.9~ms after Doppler cooling. This time we determined the
final $\bar{m}^{y,z}$ by comparing the excitation probability on the red and
the blue sidebands of the ${\rm S}_{1/2}(m=1/2) \to {\rm D}_{5/2}(m=5/2)$
transition. The result is shown in Fig.~\ref{twomodes}. We find both modes
cooled deeply inside the Lamb-Dicke regime ($\eta \sqrt{\bar{m}} \simeq 0.02
\ll 1$), with $p_0^y=58\%$ and $p_0^z=74\%$ ground state probability.

\section{EIT-cooling of linear ion strings}\label{linear}

Since EIT-cooling allows simultaneous cooling of several modes at different
frequencies, it seems to be particularly suited for ion strings in linear
traps. The frequencies of the axial vibrational eigenmodes of a linear string
have been calculated \cite{STEANE97,JAMES98} and measured \cite{NAGERL98}.
For a 10-ion string trapped in a linear trap with a center-of-mass axial
frequency of 0.7~MHz, the closest inter-ion spacing is found to be 3.0~$\mu$m
(4.5~$\mu$m for N=5). The axial vibration frequencies are 0.7~MHz, 1.22~MHz,
.. 4.6~MHz. The radial trap frequencies in a linear trap must be made
sufficiently high in order to prevent a transition from the linear
configuration of $N$ ions to a zig-zag configuration. It was estimated that
this transition occurs at $\nu_{rad}/\nu_{ax} \sim 0.73 N^{0.86}$, thus the
radial trap frequency must exceed 3.7~MHz for 10 ions (2~MHz for $N$=5)
\cite{WINELAND98}. Typically, the radial frequency is chosen higher. The
linear ion trap experiment at Innsbruck uses $\nu_{rad} \geq$ 4~MHz, the
Be$^+$ experiments at NIST \cite{TURCHETTE00} have $\nu_{rad} \geq$ 20~MHz.
Apart from the purely axial modes there exist $2N$ radial modes. Their
frequencies arrange in a band below $\nu_{rad}$ which overlaps with the axial
mode frequencies.

\begin{center}
\begin{figure}[t]
\epsfig{file=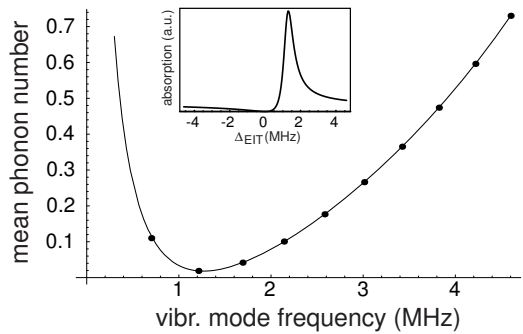, width=0.8\hsize}
\vspace{\baselineskip} \caption{EIT-cooling of the axial modes of
a linear string based on the S$_{1/2}$ - P$_{1/2}$ transition with
$\Omega_{\sigma}$=~30~MHz, $\Omega_{\pi}$=~0.5~MHz,
$\Gamma_P$=~20~MHz, and a detuning of $\Delta_{\sigma} =
\Delta_{\pi}$=~75~MHz. The axial trap frequency is 0.7~MHz. For
the calculation we have chosen the light intensity such that the
bright state is ac-Stark shifted by $\sim$ 3~MHz (see inset). The
bright resonance with a width of $\sim$ 0.5~MHz leads to cooling
for all axial modes. The mean phonon numbers (black dots) of all
axial modes are plotted versus the mode frequencies.
\label{eitlin}}
\end{figure}
\end{center}

We now estimate the performance of EIT-cooling for a 10-ion
string. The result is displayed in Fig.~\ref{eitlin}: For a 10-ion
string indeed {\em all $3N$} vibrational modes are cooled to a
mean phonon number $\bar{m}$ below one. This is promising for the
application of cold ion strings for quantum information processing
\cite{CIRAC95,MOELMER99,MOELMER00,ALMAGRO}. As discussed in the
introduction, it is required that all modes which couple to the
laser light (spectator modes) must be cooled well into the
Lamb-Dicke regime. The reason for that is that thermally excited
spectator modes cause a blurring of the Rabi frequency $\Omega$
which disturbs the precision of quantum operations. For the case
of $3N-1$ spectator modes $i$ with $\bar{m}^i$ and Lamb-Dicke
factors $\eta_i$, the initial state is a mixed state. Each time
the experiment will happen with slightly different initial
conditions and the Rabi frequency for carrier or sideband
transitions will differ. The relative blurring reads like
$\Delta\Omega / \Omega = \sqrt{\sum \eta_i^4~ \bar{m}^i~
(\bar{m}^{i}+1) / (3N-1)}$ (see equ. 126 in
ref.~\cite{WINELAND98}) which we find as small as $3 \cdot
10^{-4}$ for our specific example. The maximum number of Rabi
oscillations will then be $\sim (\Delta\Omega / \Omega)^{-1} \sim$
2000.

\section{Conclusion}

In summary we have presented EIT-cooling, a novel ground state cooling method
which uses electromagnetically induced transparency on two coupled dipole
transitions. We have demonstrated its successful application to a single
Ca$^+$ ion in a Paul trap. We have also shown several estimates for its
implementation in various trapped-ion and trapped-atom experiments, taking
into account different parameter regimes as well as deviations from the ideal
case such as extra levels and transitions. The EIT-cooling method, according
to these estimates and considering its experimental simplicity, bears the
potential to become a standard tool for the preparation of ultracold atoms
and ions. A particularly important application will be cooling of an ion
string in a linear trap, as a step in preparing such a string for coherent
manipulations in quantum information processing.

\acknowledgements This work is supported by the Austrian 'Fonds zur
F\"orderung der wissenschaftlichen Forschung' (SFB15 and START-grant
Y147-PHY), by the European Commission (TMR networks 'Quantum Information'
(ERB-FRMX-CT96-0087) and 'Quantum Structures' (ERB-FMRX-CT96-0077)), and by
the "Institut f\"ur Quanteninformation GmbH".


\end{multicols}

\begin{references}
\bibitem{noteonalpha}
$\alpha$ is between 1/5 and 2/5 depending on the type of
transition and the orientation of quantization axis and vibration,
$\alpha$=1/3 for isotropic emission, see also
\protect\cite{STENHOLM86}
\end{references}
\end{document}